\documentclass{elsart}
\usepackage{graphicx}
\usepackage{lineno}
\usepackage{upgreek} 
\usepackage{latexsym}  
\usepackage{amssymb}
\usepackage[usenames,dvipsnames]{color}

\begin{document}

\begin{frontmatter}

\title{The spectrum of high-energy cosmic rays measured with KASCADE-Grande}

\author[fzk]{W.D.~Apel},
\author[mex]{J.C.~Arteaga-Vel\'azquez\thanksref{corr}},
\author[fzk]{K.~Bekk},
\author[tor]{M.~Bertaina\thanksref{corr}},
\author[fzk,uni]{J.~Bl\"umer},
\author[fzk]{H.~Bozdog},
\author[buc]{I.M.~Brancus},
\author[sie]{P.~Buchholz},
\author[tor,toi]{E.~Cantoni},
\author[tor]{A.~Chiavassa},
\author[uni]{F.~Cossavella\thanksref{r0}},
\author[fzk]{K.~Daumiller},
\author[bra]{V.~de Souza},
\author[tor]{F.~Di~Pierro},
\author[fzk]{P.~Doll},
\author[fzk]{R.~Engel},
\author[fzk]{J.~Engler},
\author[fzk]{M. Finger}, 
\author[wup]{D.~Fuhrmann},
\author[toi]{P.L.~Ghia}, 
\author[fzk]{H.J.~Gils},
\author[wup]{R.~Glasstetter},
\author[sie]{C.~Grupen},
\author[fzk]{A.~Haungs\thanksref{corr}},
\author[fzk]{D.~Heck},
\author[nij]{J.R.~H\"orandel},
\author[uni]{D.~Huber},
\author[fzk]{T.~Huege},
\author[wup]{K.-H.~Kampert},
\author[uni]{D.~Kang\thanksref{corr}}, 
\author[sie]{D.~Kickelbick},
\author[fzk]{H.O.~Klages},
\author[uni]{K.~Link}, 
\author[pol]{P.~{\L}uczak},
\author[uni]{M.~Ludwig},
\author[fzk]{H.J.~Mathes},
\author[fzk]{H.J.~Mayer},
\author[uni]{M.~Melissas},
\author[fzk]{J.~Milke},
\author[buc]{B.~Mitrica},
\author[toi]{C.~Morello},
\author[tor]{G.~Navarra\thanksref{r2}},
\author[fzk]{J.~Oehlschl\"ager},
\author[fzk]{S.~Ostapchenko\thanksref{r3}},
\author[sie]{S.~Over},
\author[uni]{N.~Palmieri},
\author[buc]{M.~Petcu},
\author[fzk]{T.~Pierog},
\author[fzk]{H.~Rebel},
\author[fzk]{M.~Roth},
\author[fzk]{H.~Schieler},
\author[fzk]{F.G.~Schr\"oder},
\author[ubu]{O.~Sima},
\author[buc]{G.~Toma},
\author[toi]{G.C.~Trinchero},
\author[fzk]{H.~Ulrich},
\author[fzk]{A.~Weindl},
\author[fzk]{J.~Wochele},
\author[fzk]{M.~Wommer},
\author[pol]{J.~Zabierowski}

\address[fzk]{Karlsruher Institut f\"ur Technologie (KIT), Institut f\"ur Kernphysik, Germany}
\address[mex]{Universidad Michoacana, Instituto de F\'{\i}sica y Matem\'aticas, Morelia, Mexico}
\address[uni]{Karlsruher Institut f\"ur Technologie (KIT), Institut f\"ur Experimentelle Kernphysik, Germany}
\address[tor]{Dipartimento di Fisica dell' Universit\`a Torino, Italy}
\address[buc]{National Institute of Physics and Nuclear Engineering, Bucharest, Romania}
\address[sie]{Fachbereich Physik, Universit\"at Siegen, Germany}
\address[toi]{Osservatorio Astrofisico di Torino, INAF Torino, Italy}
\address[bra]{Universidade S\~ao Paulo, Instituto de F\'{\i}sica de S\~ao Carlos, Brasil}
\address[wup]{Fachbereich Physik, Universit\"at Wuppertal, Germany}
\address[nij]{Department of Astrophysics, Radboud University Nijmegen, The Netherlands}
\address[pol]{National Centre for Nuclear Research, Department of Cosmic Ray Physics, {\L}\'od\'z, Poland}
\address[ubu]{Department of Physics, University of Bucharest, Bucharest, Romania}

\thanks[corr]{corresponding authors: Juan-Carlos Arteaga-Vel\'azquez, Mario Bertaina, Donghwa Kang, 
{\it E-mail: haungs@kit.edu, spokesperson KASCADE-Grande}}
\thanks[r0]{now at: Max-Planck-Institut f\"ur Physik, M\"unchen, Germany}
\thanks[r2]{deceased}
\thanks[r3]{now at: University of Trondheim, Norway}

\begin{abstract}
The energy spectrum of cosmic rays between $10^{16}\,$eV and $10^{18}\,$eV, 
derived from measurements of the shower size (total number of charged particles)  
and the total muon number of extensive air showers by 
the KASCADE-Grande experiment, is described. 
The resulting all-particle energy spectrum exhibits strong hints for a hardening 
of the spectrum at approximately $2\cdot10^{16}\,$eV and a significant steepening 
at $\approx 8\cdot10^{16}\,$eV. 
These observations challenge the view that the spectrum is a single power law 
between knee and ankle.
Possible scenarios generating such features are discussed in terms of astrophysical 
processes that may explain the transition region from galactic to extragalactic 
origin of cosmic rays. 
\end{abstract}

\end{frontmatter}


\section{Introduction}

The main goals of experimental cosmic ray research are the determination of the arrival 
direction distribution, the primary energy spectrum, and the elemental composition. 
Those measurements comprise important hints to understand the origin, acceleration and
propagation of energetic cosmic particles. 
The needed measurements can be done directly or indirectly, depending
on the energy of the primary particle.  
At energies above $10^{15}\,$eV, the energy spectrum must be determined indirectly
from the measured properties of extensive air showers (EAS) that cosmic rays 
induce in the Earth's atmosphere~\cite{Wat}. 
 
The determination of the primary energy and elemental composition in the energy range 
from $10^{15}\,$eV up to above $10^{20}\,$eV is subject of earth-bound experiments 
since more than five decades. 
It has been shown that the all-particle spectrum has a power-law like behavior 
($\propto\,E^{-\gamma}$, with $\gamma\,\approx\,2.7$)
with features, which are known as `knee' and `ankle' at 
$3$-$5\cdot10^{15}\,$eV and $4$-$10\cdot10^{18}\,$eV, respectively. 
Whereas at the knee the spectrum steepens, the ankle is characterized by a flattening 
of the spectrum by roughly the same change of the spectral index of 
$\Delta \gamma = \pm 0.3$-$0.4$. 
Cosmic rays above the ankle are most probable of extragalactic origin~\cite{auger-eg},
i.e.~ somewhere in the energy range from $10^{16}\,$eV to a few $10^{18}\,$eV 
a break-off of the heavy component 
and the transition of cosmic rays of galactic to extragalactic origin is expected. 

As the measured position of the knee is roughly in agreement with the energy where 
supernova remnants (SNR) become inefficient accelerating particles~\cite{hillas}, 
various theories with different assumptions were developed to explain the behavior 
of the spectrum between the knee and ankle features. \\ 
The basic idea of the `dip model'~\cite{berezinsky} is that the ankle is a propagation 
feature of extragalactic protons (energy loss by electron pair production). 
Consequently, in that model the composition at the ankle is to a large extent proton-dominant 
and the transition from galactic to extragalactic origin of cosmic rays occurs already at 
energies well below $10^{18}\,$eV.   
In the scenario of the dip model, at energies around 
$10^{17}\,$eV a pure galactic iron component should be left and a small kink in the spectrum at 
around $5$-$7\cdot10^{17}\,$eV, as indicated by observations by the AKENO~\cite{akeno} and 
HiRes~\cite{hires} experiments and named as `second knee'~\cite{second}, would be assigned 
to the transition. 
This is in agreement with the SNR theory, where the knee positions of individual primary 
masses are proportional to the charge of the nuclei starting with the proton knee at 
around $E_{knee}^p = 3$-$5\cdot10^{15}\,$eV and $E_{knee}^A = Z\cdot E_{knee}^p$ 
(rigidity dependence of knee positions for galactic cosmic rays). \\
On the other hand, to avoid an early appearance of the extragalactic cosmic ray component, 
Hillas~\cite{hillas} proposed in addition to the standard SNR component, a `component B' 
of cosmic rays of galactic origin. This component would also experience a charge dependence 
of break-offs, but now shifted to approximately ten times higher energy. 
As a result, the transition occurs here at the ankle and for the entire energy range from 
$10^{15}\,$eV to $10^{18}\,$eV a mixed elemental composition is expected. 
In this scenario, the second knee, if it exists, would be a feature of the component B. 
 
The KASCADE experiment and its extension, KASCADE-Grande, aim to provide high quality 
air-shower data in the energy range of $10^{14}\,$eV to $10^{18}\,$eV to evaluate the 
validity of these models and to distinguish between them. 
The KASCADE experiment has shown that the knee is due to a distinct break 
in the proton intensity despite protons are not the most abundant primary  
in this energy range. 
The break is followed by a kink in the spectrum of Helium nuclei~\cite{kas-unf}, 
i.e.~the knee in the all-particle spectrum is a feature of the light nuclei (Z$<6$), only, 
where the difference in the energies of the knee features of primary protons and Helium facilitates 
the assumption of a charge dependence of the break-off.
First analyses of KASCADE-Grande data~\cite{kg-letter} resulted in a knee-like feature at around
$8\cdot10^{16}\,$eV caused by a steepening in the spectrum of heavy primary cosmic rays.
In the present analysis, the reconstruction of the all-particle energy spectrum of cosmic 
rays in the range from $10^{16}$ to $10^{18}\,$eV is described in detail.

 Depending on the experimental apparatus and the detection technique of ground-based 
air-shower experiments, different sets of EAS observables are available to estimate 
the energy of the primary cosmic ray~\cite{Haungs}. 
In case of ground arrays the total number of charged particles (often called shower size) 
in the shower and the corresponding particle density at observation level are 
commonly employed. 
The muon content of EAS plays an important role, too. 
In the atmosphere the muon component suffers less attenuation than electromagnetic 
or hadronic components and exhibits less fluctuations compared to the more abundant 
electromagnetic component.
In KASCADE-Grande both components, the muon and the electromagnetic ones, are measured with 
independently operating detectors. 
Both, together with the information of their correlation on a single-event-basis, are used 
to derive the spectrum. 
After a short description of the apparatus and the reconstruction procedures of the 
EAS parameters, we will describe the method developed to determine the all-particle 
energy spectrum including studies of systematic uncertainties. 
We conclude this paper with a discussion of the results.

\section{The Experiment}

The experimental layout, as well as the reconstruction procedures and accuracies of 
KASCADE-Grande observables are described in detail in reference~\cite{kg-NIM10}. 
In this chapter, we only summarize the most important facts relevant for the present analysis. 

\subsection{KASCADE-Grande}

\begin{figure}[t]
\vspace*{0.05cm}
\begin{center}
\includegraphics [width=0.6\textwidth]{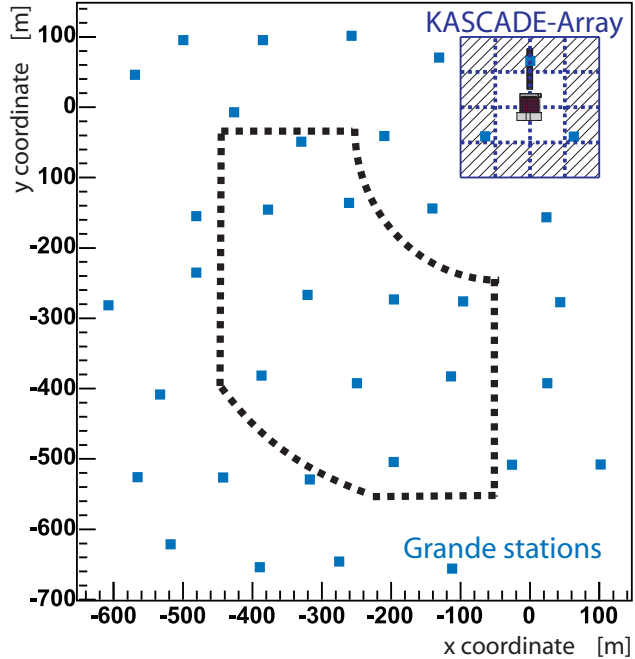}
\end{center}
\vspace*{-0.1cm}
\caption{Layout of the KASCADE-Grande experiment:
The KASCADE array and the distribution of the 37 stations of the Grande
array are shown. The outer 12 clusters of the KASCADE array consist of 
shielded $\mu$-detectors (hatched area). The dotted line shows the fiducial area 
selected for the present analysis.}
\label{fig1}
\vspace*{-0.01cm}
\end{figure}

The multi-detector experiment KASCADE~\cite{kascade}
(located at 49.1$^\circ$N, 8.4$^\circ$E, 110$\,$m$\,$a.s.l.)
was extended to KASCADE-Grande 
in 2003 by installing a large array of 37 stations consisting 
of 10$\,$m$^2$ scintillation detectors each (Fig.~\ref{fig1}).  
KASCADE-Grande provides a sensitive area of about $0.5\,$km$^2$
and operates jointly with the existing KASCADE detectors.
Main parts of the experiment used for the present analysis are the Grande array
spread over an area of $700 \times 700\,$m$^2$, and the original KASCADE array 
covering $200 \times 200\,$m$^2$. 
The Grande array is installed over an irregular 
triangular grid with an average spacing of $137\,$m. 
The KASCADE array is composed of 252 detector stations on a square grid with $13\,$m spacing. 
It is organized in 12 outer clusters of 16 stations each and 4 inner clusters of 15 stations each. 
The outer clusters (192 stations) are equipped with two unshielded (e.m.)
and one shielded (muon) detector units each. 
A muon detector unit consists of 4 plastic scintillators of $90 \times 90 \times 3$\,cm$^3$ each,
where the iron-lead shielding provides a  
threshold of 230 MeV kinetic energy for vertically incident muons. 
The total sensitive area of the muon array amounts to $622\,$m$^2$. \\
While with the Grande array we reconstruct the total number of charged particles, 
data from the shielded scintillation detectors of the KASCADE array are used to 
reconstruct the total number of muons on an event-by-event basis for events 
triggered by Grande. 
The combination of both allows us to reconstruct the energy spectrum 
of cosmic rays in the range from $10^{16}\,$eV up to $10^{18}\,$eV.

\subsection{Monte Carlo simulations}

The simulations performed include the detailed air shower 
development in the atmosphere, as well as the response of the detector and its electronics. 
Therefore, the EAS parameters reconstructed 
from simulated showers are obtained exactly in the same way as for real data. 
The EAS were generated with uniformly distributed core positions (at an area larger than 
the Grande array), with isotropically distributed arrival directions, and with a 
spectral index of $\gamma=-2$, i.e. roughly one order of magnitude harder than the 
measured spectrum. 
The spectral slope is chosen as a compromise between computing time and 
statistical accuracy at the highest energies.
Later, in the analysis procedures the simulated data is weighted to describe a 
softer energy spectrum with $\gamma=-3$.
The EAS were simulated with CORSIKA~\cite{cors} and the
Monte Carlo generators FLUKA~\cite{fluka} and QGSJet~II~\cite{qgs} (hadronic interactions). 
Sets of simulated events were produced in the energy range from 
$10^{15}\,$eV to $10^{18}\,$eV for five different representative mass groups: 
H, He, C, Si and Fe with about $353,000$ events per primary.
A few showers for all primaries were also simulated for the higher energy range up 
to $3 \cdot 10^{18}\,$eV in order to study the reconstruction quality even beyond 
the energy range of interest. 
In addition, with less statistics, similar simulation sets were generated based 
on the high-energy hadronic interaction model EPOS, version~1.99~\cite{epos}.  
 
\subsection{Reconstruction}

Basic shower observables like the core position and the angle-of-incidence, 
as well as the total number of charged particles are provided by 
the reconstruction of data recorded by the Grande stations. 
The angle-of-incidence is determined using the relative particle arrival 
times at the stations. 
The core location, the slope of the lateral distribution function and the shower size 
(i.e.~the total number of charged particles $N_{ch}$) are calculated by means 
of a maximum likelihood procedure, comparing the measured number of particles 
with the one expected from a NKG-like lateral distribution function~\cite{hajo} 
of charged particles in the EAS.  
KASCADE-Grande provides the unique opportunity of evaluating the reconstruction 
accuracies of the Grande array by a direct comparison with an independent experiment.
For a subsample of events collected by the Grande array it is possible to 
compare on an event-by-event basis the two independent reconstructions of KASCADE and
Grande. By means of such a comparison the Grande reconstruction accuracies of the total 
number of charged particles are found to be: 
systematic uncertainty $\le 5$\%, statistical accuracy better than 
$15$\%; accuracy of the arrival direction of about $0.8^\circ$; accuracy
of the core position about $6\,$m. 
The total number of muons $N_\upmu$ is calculated using the core position 
determined by the Grande array and the muon densities ($E_\upmu > 230\,$MeV) 
measured by the KASCADE muon array detectors. 
$N_\upmu$ is derived from a maximum likelihood estimation comparing the measured 
densities with a lateral distribution function, where the uncertainty is less than 12\%.
For the purpose of the present analysis, the estimated $N_\upmu$ has been corrected for 
systematic uncertainties using a correction function, which is derived from Monte Carlo
simulations as explained in appendix~\ref{appmuon}. 
The correction improves the accuracy to better than 10\%, including the  
uncertainty due to hadronic interaction models and unknown composition.

\subsection{Data selection and shower size spectra}

The selection used for the present analysis requires that the events passed 
successfully the full KASCADE-Grande reconstruction procedure. 
In addition, we requested events to be from stable periods of data taking with more 
than 35 Grande stations and all 16 KASCADE clusters in operation. 
The selected events had to pass a 7/7 Grande hardware trigger (six of a hexagonal 
shape and the central one) and had to hit more than 11 Grande stations in total. \\
We restricted ourselves to events with zenith angles 
less than $40^\circ$ due to increasing shower size uncertainties. 
Only air showers with cores located in a central area of KASCADE-Grande (see 
Fig.~\ref{fig1}) were selected. With this cut on the fiducial area, 
border effects are discarded and the under- and overestimations on the muon number 
for events close to and far away from the center of the KASCADE array are reduced. 
All of these cuts were applied also to the Monte Carlo simulated events to study 
their effects on the selection, to optimize the cuts, and to control 
the uncertainty of the acceptance.
Full efficiency for triggering and reconstructing air-showers, as well as 
a uniform distribution of the shower cores in the fiducial area, is reached   
at a primary energy of about $10^{16}\,$eV, slightly depending on the 
zenith angle, and on the primary particle type. 
\begin{figure}[t]
\vspace*{0.05cm}
\begin{center}
\includegraphics [width=0.68\textwidth]{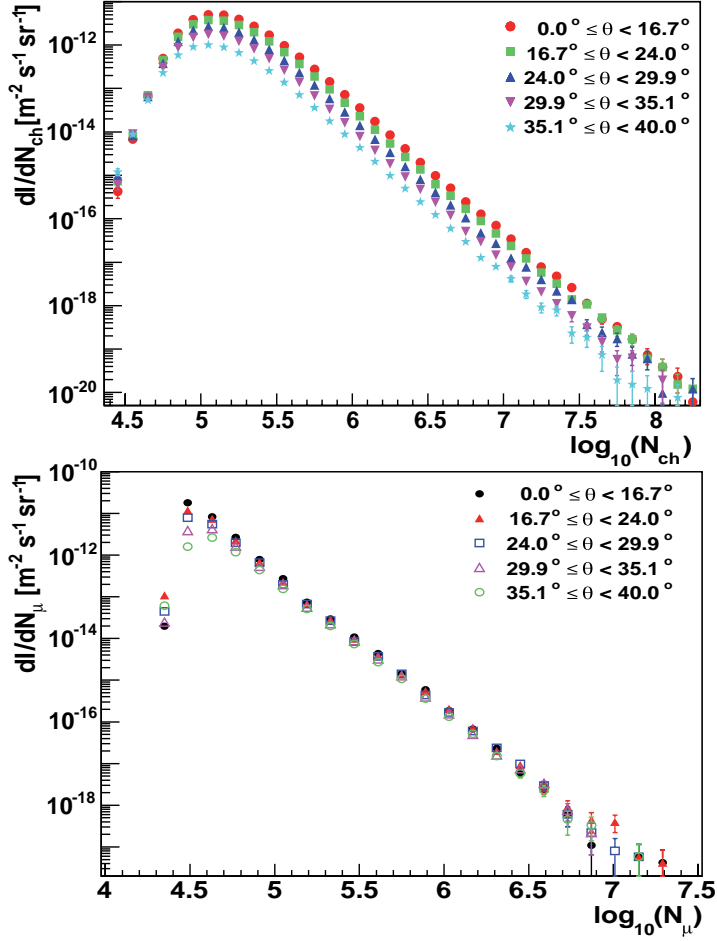}
\end{center}
\vspace*{-0.1cm}
\caption{Differential shower size $N_{ch}$ and muon number $N_{\upmu}$ spectra for different 
zenith angular ranges.}
\label{figrawspecs}
\vspace*{-0.01cm}
\end{figure}

The analysis presented here is based on 1173 days of data. The cuts on the 
fiducial area and zenith angle result in a total acceptance of 
$1.98 \cdot 10^{5}\,$m$^2 \cdot$sr, and an exposure of 
$2.0 \cdot 10^{13}\,$m$^2 \cdot$s$ \cdot$sr, respectively. 
Approximately $1.5 \cdot 10^6$ events are subject of the following analysis. \\
Fig.~\ref{figrawspecs} shows the reconstructed $N_{ch}$ and $N_\upmu$ distributions 
for five different angular ranges. 
The angular ranges were chosen in order to have the same acceptance in each of them.

\section{The reconstruction of the all-particle energy spectrum}

KASCADE-Grande, with the possibility of an independent reconstruction of the two 
observables $N_{ch}$ and $N_\upmu$ per individual EAS, allows us to go for a dedicated
strategy in estimating the all-particle energy spectrum. 
By means of Monte Carlo simulations, calibration formulas are obtained 
to calculate the primary energy per individual event based on $N_{ch}$ and $N_\upmu$, 
taking into account the correlation between the two particle components. 
This reduces the composition dependence of the energy assignment. 
To account for attenuation effects in the atmosphere, which are different for the two shower 
observables, the energy calibration is performed separately for 
the five zenith angular ranges. 
Finally, the obtained energy spectra are unfolded to account for the bin-to-bin migrations 
before they are combined to the resulting all-particle energy spectrum. 

\subsection{Energy calibration}

\begin{figure}[b]
\vspace*{0.05cm}
\begin{center}
\includegraphics [width=0.98\textwidth]{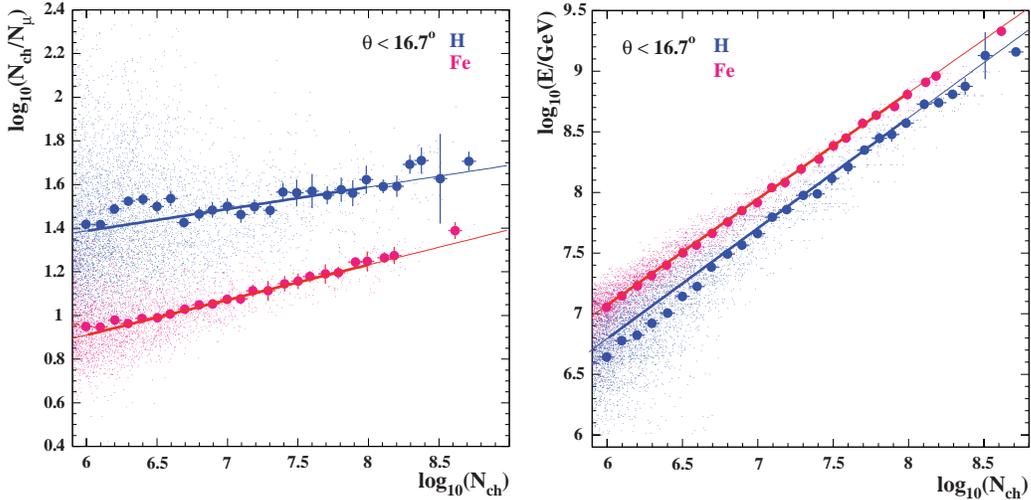}
\end{center}
\vspace*{-0.1cm}
\caption{Left panel: Scatter plot of the reconstructed $N_{ch}/N_{\upmu}$ vs. $N_{ch}$ 
for primary iron and proton nuclei and for the first angular bin. 
The full dots and error bars indicate the mean and statistical errors of the distribution 
of the individual events (small dots). The fits result in parameters $c$
and $d$ of expression~\ref{equn3}. 
Right panel: Scatter plots of E vs. $N_{ch}$ for iron and proton 
primary nuclei. The fits result in parameters $a$
and $b$ of expression~\ref{equn1}.}
\label{figfits}
\vspace*{-0.01cm}
\end{figure}
The energy assignment starts by applying $E = f(N_{ch}$$,k)$, where $k$ is defined 
through the ratio of  N$_{ch}$ and  N$_\upmu$: $ k = g (N_{ch}$,$N_\upmu)$.  
The main aim of the $k$ parameter is to correlate these observables on an
event-by-event basis, taking into account the differences in the N$_{ch}/$N$_\upmu$ ratio for 
different primary masses with the same N$_{ch}$, as well as the shower-to-shower 
fluctuations for events of the same primary mass: 

{ 
  \begin{eqnarray}
  log_{10}(E/GeV)&{}={}&[a_H+(a_{Fe}-a_H) \cdot k] \cdot log_{10}(N_{ch})+\nonumber\\
  &&{+}\: b_H+(b_{Fe}-b_H) \cdot k
  \label{equn1}
  \end{eqnarray}              %
  \begin{equation}
    k = \frac{log_{10}(N_{ch}/N_{\mu})-log_{10}(N_{ch}/N_{\mu})_H}
        {log_{10}(N_{ch}/N_{\mu})_{Fe}-log_{10}(N_{ch}/N_{\mu})_H}
    \label{equn2}
   \end{equation}
  \begin{equation}
    log_{10}(N_{ch}/N_{\mu})_{H,Fe} = c_{H,Fe} \cdot log_{10} (N_{ch}) + d_{H,Fe}.
    \label{equn3}
   \end{equation}
}
By definition the $k$ parameter is a number centered around zero for proton 
initiated showers and around one for iron initiated showers~\cite{bertaina}.
The coefficients $a, b, c, d$ are obtained independently by simulations for each zenith angular 
range and for each primary mass, where fits are applied to the scatter plots 
$(N_{ch},N_{ch}/N_{\upmu})$ and $(N_{ch},E)$.  
The fit range is chosen to be $6 \le \log(N_{ch}) \le 8$, i.e.~where 100\% trigger 
efficiency is guaranteed.
Primary protons exhibit larger fluctuations than heavier primaries, therefore, the coefficients 
$c$ and $d$ are obtained iteratively in case of protons in order to improve the reconstruction 
of the energy spectrum.  
As an example, Fig.~\ref{figfits} shows the scatter plots including the resulting functions 
for the first angular bin. Shown are the errors on the mean, which are small due to the large 
Monte Carlo statistics. For the fits, however, we also take into account the width of the distributions 
in order to avoid a bias due to varying shower-to-shower fluctuations, in particular in case of 
primary protons and small shower sizes.
It is obvious that taking into account the correlation of the observables reduces significantly 
the composition dependence of the energy assignment.      
Similar procedures are applied to the other angular bins and all the coefficients are compiled 
in Table~\ref{Tab01}. The uncertainties of these numbers are small, but considered in the calculation of 
the total systematic uncertainty. \\
 \begin{table*}[h]
  \caption{Coefficients of the energy calibration functions.}
  \vspace{0.5pc}
   {\scriptsize
  \centering
  \begin{tabular}{|l|c|c|c|c|c|c|c|c|}
  \hline
   Angular bin & \multicolumn{2}{c|}{a} & \multicolumn{2}{c|}{b}& \multicolumn{2}{c|}{c}& \multicolumn{2}{c|}{d} \\ \hline
                                              &   H   & Fe    &   H   & Fe    &   H   & Fe    &   H   & Fe     \\ \hline \hline
         $\theta < 16.7^\circ$                & 0.91 & 0.88 & 1.33 & 1.82 & 0.10 & 0.16 & 0.79 & -0.06 \\ \hline
         $16.7^\circ \le \theta < 24.0^\circ$ & 0.89 & 0.88 & 1.50 & 1.92 & 0.08 & 0.18 & 0.88 & -0.25 \\ \hline
         $24.0^\circ \le \theta < 29.9^\circ$ & 0.94 & 0.89 & 1.30 & 1.94 & 0.10 & 0.16 & 0.68 & -0.17 \\ \hline
         $29.9^\circ \le \theta < 35.1^\circ$ & 0.93 & 0.88 & 1.46 & 2.10 & 0.11 & 0.17 & 0.54 & -0.35 \\ \hline
         $35.1^\circ \le \theta < 40.0^\circ$ & 0.92 & 0.88 & 1.75 & 2.29 & 0.11 & 0.16 & 0.41 & -0.35 \\ \hline
  \end{tabular}
  \label{Tab01}
   }
 \end{table*}   

\begin{figure}[t]
\vspace*{0.05cm}
\begin{center}
\includegraphics [width=0.6\textwidth]{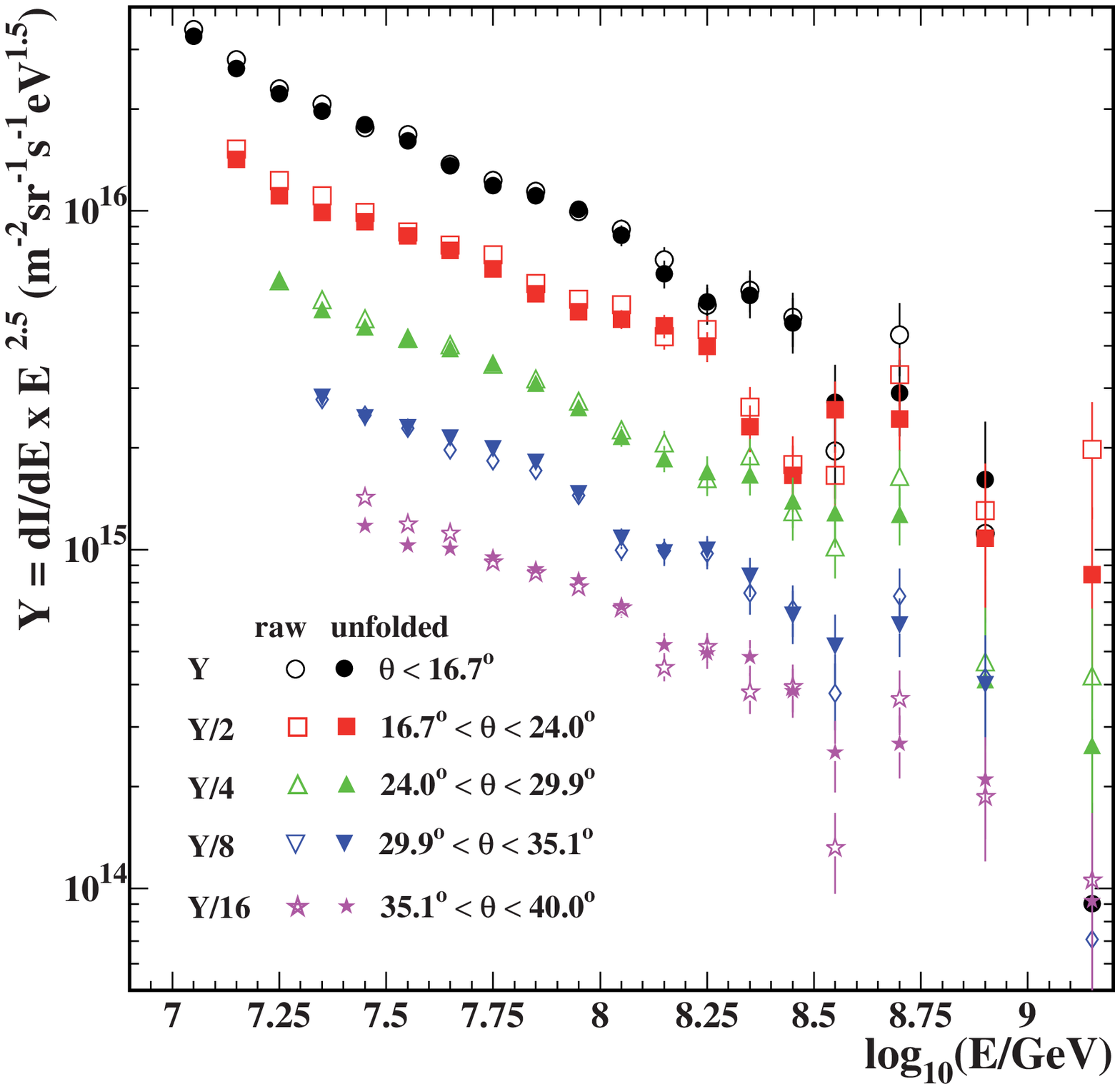}
\includegraphics [width=0.6\textwidth]{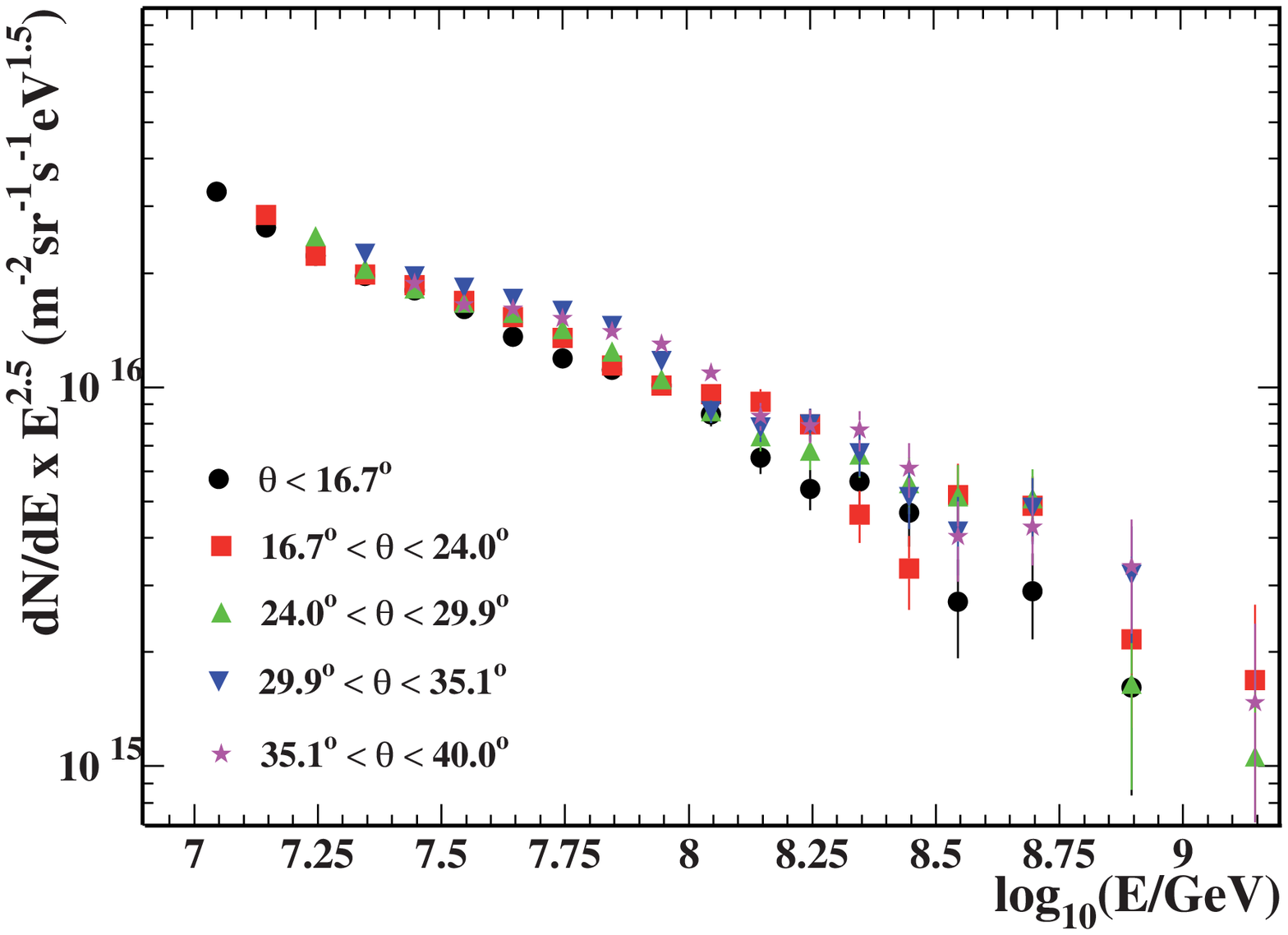}
\end{center}
\vspace*{-0.3cm}
\caption{Reconstructed all-particle energy spectra for all five angular bins. 
In the upper panel the direct reconstructed as well as the unfolded spectra are displayed, where the 
spectra are scaled for better visibility. 
In the lower panel only the unfolded spectra are shown without scaling.
Only statistical uncertainties are displayed.}
\label{mariospecs}
\vspace*{0.01cm}
\end{figure}

Applying the derived energy calibration to the measured data, all-particle 
energy spectra for the five zenith angular ranges are obtained (Fig.~\ref{mariospecs}).
To refine the energy assignment function from the so far assumed 
pure power-law behavior of the $(N_{ch},N_{ch}/N_{\upmu})$ and $(N_{ch},E)$ relations to a more 
realistic non-linear calibration, as well as to unfold bin-to-bin migrations due to 
shower-to-shower fluctuations,
response matrices $R_{ij}$ for the different angular bins are constructed and applied, i.e. the 
spectra are unfolded (see appendix~\ref{appunf}). 
Effects of this procedure (see Fig.~\ref{mariospecs}) on the flux are estimated to be smaller 
than 5\% at all energy bins and, therefore, do not significantly change the shape of the spectra. 
For the following discussions we always refer to the unfolded spectra.

The spectra of the different angular ranges exhibit small systematic shifts relative to each other
(see Fig.~\ref{mariospecs}), where we observe a slight flux increase with increasing zenith angle. 
This corresponds to a horizontal shift in the energy assignment, which could be explained if
real showers penetrate deeper in the atmosphere than predicted by the QGSJetII hadronic interaction 
model\footnote{In fact, the analysis of muon production heights with the muon tracking detector of 
KASCADE leads to a similar conclusion, see ref.~\cite{MTD}.}.
As the effect is observed for different assumed composition models and for all hadronic interaction models,
it is most probably caused by a mismatch between the predicted and measured 
attenuation lengths of the shower particle numbers $N_{ch}$ and $N_\upmu$ 
(for a more detailed discussion see appendix~\ref{appatt}). 
The aforementioned differences in the spectra are considered as one of the major sources of 
the systematic uncertainty on the energy spectrum, and are taken into account in the estimation 
of the total systematic uncertainties.
The final all-particle spectrum of KASCADE-Grande is obtained 
(see Figs.~\ref{specscomp},~\ref{residual},~\ref{spectrum}, and Table~\ref{Tab02}) 
by combining the spectra for the individual angular ranges.  
Only those events are taken into account,
for which the reconstructed energy is above the energy threshold for 
the angular bin of interest (see Fig.~\ref{mariospecs}). 

\subsection{Systematic uncertainties}

Different sources of systematic uncertainties, which affect 
the all-particle energy spectrum, are investigated. Most of the effects lead to a shift 
in energy and this, in turn, to a shift in the spectrum (see also Table~\ref{tab-uncertainty}):  
\begin{itemize}
\item Attenuation:
The average difference between the intensities obtained for the various angular bins has been used to 
define the systematic uncertainty associated with the
angular dependence of the parameters appearing in the energy calibration functions of 
the different angular ranges.
This includes the systematic uncertainty related to the description of the air shower 
attenuation in the atmosphere for simulated data. 
The uncertainty is used as well at lower energies, where not all angular bins contribute to the spectrum. 
In addition, this uncertainty indicates the limit for which the QGSjetII model reproduces the 
shower attenuation in a consistent way for the selected data sample. 
 \begin{table*}[ht]
  \caption{Estimated uncertainties of the cosmic ray intensity for different energies, 
  where only absolute values are given in case of symmetric uncertainties.}
  \vspace{0.5pc}
   {\scriptsize
  \centering
  \begin{tabular}{|l|c|c|c|}
  \hline
   Source of uncertainty & 10$^{16}$eV (\%) & 10$^{17}$eV (\%) & 10$^{18}$eV (\%) \\ \hline \hline
   intensity in different angular bins (attenuation)& $-0$/$+6.5$ & 10.9 & 21.3\\ \hline 
   energy calibration and composition & 10.3 & 5.8 & 13.4 \\ \hline   
   slope of the primary spectrum & 4.0 & 2.0 & 1.9 \\ \hline  
   reconstruction (core and shower sizes) & 0.1 & 1.4 & 6.5 \\ \hline 
   total & -11.1/+12.8 & 12.6 & 26.1\\ \hline \hline 
   artificial spectrum structures (extreme cases) &  & $<$10 &  \\ \hline 
   hadronic interaction model (EPOS-QGSJet) & -5.3 & -16.9 & -14.6 \\ \hline 
   statistical error & 0.6 & 2.7 & 17.0 \\ \hline \hline
   energy resolution (mixed composition) & 24.7 & 18.6 & 13.6 \\ \hline
  \end{tabular}
  \label{tab-uncertainty}
   }
 \end{table*}   
\begin{figure}[t]
\vspace*{0.05cm}
\begin{center}
\includegraphics [width=0.98\textwidth]{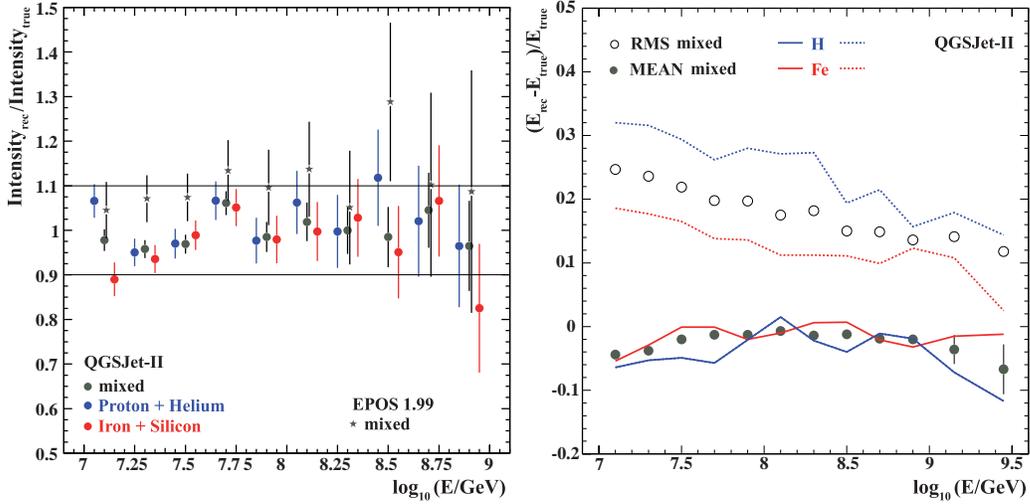}
\end{center}
\vspace*{-0.1cm}
\caption{Left panel: Ratio between the reconstructed and initially 
simulated energy spectrum  for light (blue), heavy (red) and mixed
primaries (black) summing up all angular bins.
The ratio using simulated EPOS data when the spectrum is reconstructed 
using the QGSJet-II based calibration functions is also shown.
The larger fluctuations are due to the lower statistics of the available EPOS simulations.
Right panel: Resolution of the energy assignment for  simulated sets (QGSJet) with mixed 
composition, pure proton, and pure iron primaries, respectively. The dots show the
offsets of the reconstructed energy in bins of simulated energy for mixed 
composition, while lines represent the results for pure H and Fe assumptions. 
The circles and dotted lines show the corresponding RMS of these distributions.}
\label{figcompare}
\vspace*{-0.01cm}
\end{figure}
\item Energy calibration and composition: 
To estimate a possible bias in reproducing the energy spectrum, eqns.~\ref{equn1} and~\ref{equn2} 
have been applied to simulated energy spectra build up by pure H and Fe primaries as well as the 
other three mass groups (He, C, Si), and by a uniform mixture of the five primaries with 
20\% abundance each. 
The spectra have been composed to fall off with slope $\gamma$ = -3 and are compared with the 
reconstructed ones to find the corresponding uncertainty in intensity. 
Figure~\ref{figcompare} summarizes
the results on the ratio between the reconstructed and the initally simulated spectra 
for the full angular range. 
The original energy spectra are fairly well reproduced in all energy bins within a systematic 
uncertainty smaller than 10$\%$. 
This is quite important because it guarantees that any primary spectrum with arbitrary composition 
can be fairly well reproduced, even though the parameters of the response matrix and of the 
muon number correction functions have been defined on the basis of a mixed composition.
Another important aspect to investigate is the capability and robustness of the analysis 
to recognize changes of the spectral shape and the elemental composition in the considered spectrum. 
Different cases have been simulated, from rigidity-dependent knees with variable 
abundances for each single chemical group to completely artificial compositions, 
such as sudden changes around 10$^{17}\,$eV from nearly pure iron to 
nearly pure protons and vice versa. 
Even for these extreme cases, the ratio between the reconstructed and initially simulated intensity
deviates from unity throughout by less than 10$\%$.
\item Spectral slope of the Monte Carlo simulations: 
A further source of systematic uncertainty is the choice of a spectral slope of $\gamma$ = -3 
in the simulations to determine the energy calibration functions and response matrices. 
Therefore, new calibration functions and response matrices have been calculated using 
simulated spectra with $\gamma_1$ = -2.8 and $\gamma_2$ = -3.2. 
The difference between the intensities obtained with the new coefficients has been defined 
as systematic uncertainty and results into 4\% at low energies, where the fluctuations are 
more important, and decreases to about 2\% at E$\approx$10$^{17}$ eV.
\item Reconstruction quality of shower sizes: 
The Grande array has an asymmetric geometry. Therefore, we checked the reconstruction of 
the observables by determining the energy spectrum for different ranges of distance of 
the shower core to the muon detector. 
The relative difference in intensity as a function of energy is used to compute a 
systematic uncertainty induced by the reconstruction, and it amounts to about $2$\% 
at around $10^{17}\,$eV, slightly increasing with energy.
\item Energy resolution: 
Simulated data using an equal mixture of all primaries have been divided in bins of true energy 
and the distributions of the relative differences between reconstructed and true energies 
have been compared. As shown in Fig.~\ref{figcompare}, right panel, 
the RMS of such distributions (energy resolution) is about $25$\% at lower energies and 
decreases to about $15$\% at the highest energies due to the decrease of intrinsic shower 
fluctuations. 
Results for pure H and Fe primaries are also indicated by lines. 
As expected, proton initiated showers show larger fluctuations compared to EAS generated by 
primary iron nuclei.
\item Hadronic interaction models: 
By now, for all considerations the model combination QGSJet-II/FLUKA has been used. 
As the calibration depends on simulations, other interaction models may
change the interpretation of the data. To study such effects we investigated the influence of 
the hadronic interaction model by performing the energy assignment based on 
simulations with the hadronic interaction model EPOS (version 1.99). Due to the smaller 
statistics of the EPOS simulations, larger uncertainties are obtained and no response matrix 
corrections could be applied. 
Therefore, a proper energy spectrum by means of EPOS simulations cannot be derived yet, but 
general characteristics are inferred. Comparing the all-particle energy spectra for both cases,
it was found that EPOS leads to a slightly lower intensity ($\approx$10-15\%) compared to QGSJet. 
A similar conclusion can be drawn, if the data simulated with EPOS are treated like experimental data 
and reconstructed using the QGSJet based calibration functions (see Fig.~\ref{figcompare}). 
Now, as a consequence, the intensity is reconstructed 10-15\% higher than the simulated input. 
The main reason behind the difference between QGSJet and EPOS results 
has to be ascribed to the different $N_{ch}/N_\upmu$ ratio predicted by the two models. 
In particular, EPOS predicts that the showers are richer in muons and slightly poorer in charged 
particles (for the relevant energy range, at sea level, and for the experimental conditions of 
KASCADE-Grande). These results are used as a rough estimate of the uncertainties due to the choice  
of the particular interaction model QGSJet-II. 
\end{itemize}

\subsection{Cross-checks of the energy spectrum reconstruction} 

Using the $N_{ch}$-$N_{\upmu}$ ratio we reduced the dependence of the reconstructed 
all-particle spectrum on the elemental composition. 
But, since both observables are reconstructed independently, we can apply an energy spectrum 
reconstruction on both observables individually.
At the end one would expect the same result for the energy spectrum by all approaches,
provided that (i) the measurements are accurate enough, (ii) the reconstructions work 
without failures, and (iii) the Monte Carlo simulations describe correctly the shower 
development and its fluctuations, and (iv) the composition is known. 
But, the fact that not all of the above requirements are fulfilled and the individual 
observables exhibit substantial differences in their  
composition sensitivity hampers such straightforward cross-checks.  
Nevertheless, such analyses are used to check the reconstruction procedures and the  
influence of systematic uncertainties.
Some details of these analyses~\cite{kang,arteaga} and the results can be found in 
appendix~\ref{apponed}. 
 
In figure~\ref{specscomp} the spectra obtained by the three methods are compiled, 
where the intensity is multiplied by a factor of $E^{3.0}$. 
Owing to the different approaches, the results using single observables
are given only for the pure proton and iron assumptions, whereas the final spectrum
is displayed with a band showing the systematic uncertainties. 
Using $N_{ch}$ as the observable obtained from data of the Grande array only, a large 
dependence on the primary elemental composition is present, which is reflected in a big  
difference between the intensities for proton and iron assumptions. 
The muon number $N_{\upmu}$ shows less composition dependence compared to the shower size, though
it is still the largest contribution of uncertainty.
The narrower range for a solution provided by $N_\upmu$ compared to $N_{ch}$ 
confirms the finding of KASCADE that at sea-level the number of mostly low-energy muons 
$N_{\upmu}$ is less composition sensitive than the total 
number of charged particles~\cite{KAS-roth}. 
The method finally applied for energy estimation, owing to the combination of the two variables, 
results in a larger reconstruction uncertainty. But the total uncertainty (including composition dependence) 
is considerably smaller by taking into account also the correlation of these observables. This 
additional information is strikingly decreasing the composition dependence. 
\begin{figure}[t]
\vspace*{0.05cm}
\begin{center}
\includegraphics [width=0.68\textwidth]{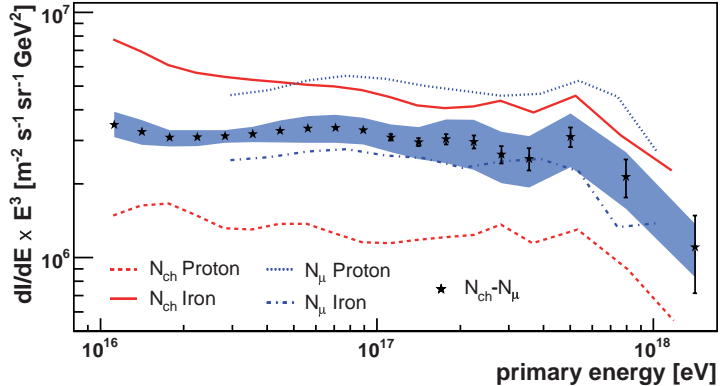}
\end{center}
\vspace*{-0.3cm}
\caption{Reconstructed all-particle energy spectrum by the three different approaches
applied to KASCADE-Grande data. For the combined method a band is shown, indicating the 
range of the systematic uncertainty (without uncertainties due to the chosen 
hadronic interaction model).}
\label{specscomp}
\vspace*{0.01cm}
\end{figure}

Of particular interest is the fact that by using $N_{ch}$, the iron assumption results in a
higher intensity than the proton assumption, whereas using $N_{\upmu}$ the opposite is the 
case (see appendix~\ref{apponed}). 
In any case, if there is only the possibility of applying a one-dimensional method, then one finds a 
large variance in possible solutions (any solution within the range spanned by the 
proton and iron line, not even parallel to these lines). 
Interestingly, over the whole energy range there is only little room for a solution satisfying both 
ranges, spanned by $N_{ch}$ and $N_{\upmu}$, and this solution prefers a more heavy 
composition - in the framework of the QGSJet-II hadronic interaction model.  
The results obtained by combining $N_{ch}$ and $N_\upmu$ lie within
the area spanned by the other methods. This finding expresses a cross-check of the 
intrinsic consistency of the results for the interpretation of two measured observables based on the used 
hadronic interaction model QGSJet-II. This was found to be valid also for 
the case of interpreting the data with the hadronic interaction model EPOS-1.99.

\section{Results and discussion}

\begin{figure}[t]
\vspace*{0.05cm}
\begin{center}
\includegraphics [width=0.68\textwidth]{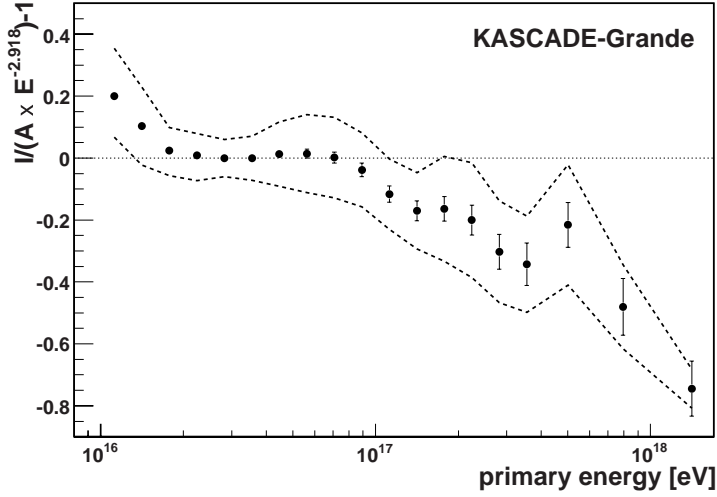}
\end{center}
\vspace*{-0.3cm}
\caption{The all-particle energy spectrum obtained with
KASCADE-Grande. The residual intensity after multiplying the spectrum 
with a factor of $E^{2.918}$ and normalized with $A$ is displayed as 
well as the band of systematic uncertainty.}
\label{residual}
\vspace*{0.01cm}
\end{figure}

The resulting all-particle spectrum exhibits structures 
which do not allow us to describe the spectrum with a single power law. 
To emphasize this, figure~\ref{residual} shows the residuals of the all-particle energy 
spectrum multiplied by a factor in such a way that the middle part of the spectrum becomes flat. 
The power law index of $\gamma=-2.92 \pm 0.02$ is obtained by fitting the range of 
$\log_{10} (E/eV) = 16.2 - 17.0$. 
For the full energy range a statistical analysis reveals that the spectrum is not described
by a single power law with a significance of $2.1 \sigma$. 
\begin{table*}[t]
  \caption{Differential intensity values of the all particle energy spectrum for the QGSJet-II based analysis. 
  The first column of errors denotes the statistical uncertainty, the second column the systematic uncertainty.}
  \vspace{0.5pc}
   {\scriptsize
  \centering
  \begin{tabular}{llc}
  \hline
   bin number & Energy [eV] & $dI/dE \pm$ stat. $\pm$ syst. \\ 
   						&	            & $[m^{-2} s^{-1} sr^{-1} GeV^{-1}]$ \\
  \hline
	1    &    $1.11 \cdot 10^{16}$  & $(2.46 \pm 0.02 \pm ^{0.27}_{0.32}) \cdot 10^{-15}$  \\  
	2    &    $1.41 \cdot 10^{16}$  & $(1.16 \pm 0.01 \pm ^{0.13}_{0.13}) \cdot 10^{-15}$  \\  
	3    &    $1.78 \cdot 10^{16}$  & $(5.49 \pm 0.03 \pm ^{0.43}_{0.40}) \cdot 10^{-16}$  \\  
	4    &    $2.24 \cdot 10^{16}$  & $(2.76 \pm 0.02 \pm ^{0.22}_{0.19}) \cdot 10^{-16}$  \\  
	5    &    $2.82 \cdot 10^{16}$  & $(1.40 \pm 0.01 \pm 0.08) \cdot 10^{-16}$  \\  
	6    &    $3.55 \cdot 10^{16}$  & $(7.13 \pm 0.07 \pm 0.50) \cdot 10^{-17}$  \\  
	7    &    $4.47 \cdot 10^{16}$  & $(3.69 \pm 0.04 \pm 0.38) \cdot 10^{-17}$  \\  
	8    &    $5.62 \cdot 10^{16}$  & $(1.89 \pm 0.03 \pm 0.23) \cdot 10^{-17}$  \\  
	9    &    $7.08 \cdot 10^{16}$  & $(9.52 \pm 0.17 \pm 1.24) \cdot 10^{-18}$  \\  
	10   &    $8.91 \cdot 10^{16}$  & $(4.67 \pm 0.11 \pm 0.58) \cdot 10^{-18}$  \\  
	11   &    $1.12 \cdot 10^{17}$  & $(2.19 \pm 0.07 \pm 0.28) \cdot 10^{-18}$  \\  
	12   &    $1.41 \cdot 10^{17}$  & $(1.05 \pm 0.04 \pm 0.16) \cdot 10^{-18}$  \\  
	13   &    $1.78 \cdot 10^{17}$  & $(5.41 \pm 0.26 \pm 1.10) \cdot 10^{-19}$  \\  
	14   &    $2.24 \cdot 10^{17}$  & $(2.64 \pm 0.16 \pm 0.61) \cdot 10^{-19}$  \\  
	15   &    $2.82 \cdot 10^{17}$  & $(1.12 \pm 0.10 \pm 0.28) \cdot 10^{-19}$  \\  
	16   &    $3.55 \cdot 10^{17}$  & $(5.66 \pm 0.59 \pm 1.34) \cdot 10^{-20}$  \\  
	17   &    $5.01 \cdot 10^{17}$  & $(2.47 \pm 0.23 \pm 0.61) \cdot 10^{-20}$  \\  
	18   &    $7.94 \cdot 10^{17}$  & $(4.26 \pm 0.75 \pm 1.11) \cdot 10^{-21}$  \\  
	19   &    $1.41 \cdot 10^{18}$  & $(3.90 \pm 1.36 \pm 0.96) \cdot 10^{-22}$  \\  
  \hline
  \end{tabular}
  \label{Tab02}
  }
 \end{table*}   

Just above $10^{16}\,$eV the spectrum exhibits a `concave' behavior, 
which is significant with respect to the systematic and statistical uncertainties.    
This is true despite the fact that only vertical showers contribute to the spectrum in 
this energy range (see Fig.~\ref{mariospecs}). 
This hardening of the spectrum is validated by several cross-checks, 
e.g., by efficiency correction of more inclined events based on simulations. 
A hardening of the spectrum is, on the one hand, expected when a pure rigidity dependence of the 
galactic cosmic rays is assumed. Depending on the relative abundances of the different primaries 
one would expect charge dependent steps (i.e.~slope changes) in the all-particle spectrum.    
The gap in the knee positions of light primaries (proton, helium, and CNO group 
of $Z=1-8$) and the heavy group can lead to a hardening of the spectrum~\cite{donato}. 
On the other hand, there are also other possible astrophysical scenarios to get a concave behavior of 
the cosmic ray spectrum. In general, a transition from one source population to another one could 
also result in a hardening of the spectrum. In such a case, the KASCADE-Grande result could be a first 
experimental hint to the `component B' of galactic cosmic rays, as proposed by Hillas~\cite{hillas}.
A possible scenario for the component B is discussed by Ptuskin et al.~\cite{Ptuskin}, where
the maximum acceleration energy for different types of supernovae is considered, taking into 
account also their relative abundances in our galaxy. This scenario can lead to an extension 
of the galactic component up to a few EeV.
In addition, in this model the transition of CR origin from the standard type SN Ia to SN IIb 
supernovae requires a hardening of the spectrum at 10 PeV. 
General galactic modulation can lead only to a very smooth change of the slope index over more 
than a decade in energy, but postulating a contribution of a nearby (single)
source, sharpening the knee at a few PeV~\cite{erlykin,shibata}, would also require a hardening 
of the spectrum just above $10\,$PeV.
It is interesting to note that recently the CREAM detector (balloon experiment) 
has described such a hardening of the proton and helium spectra at much 
lower energies~\cite{cream}, which by the authors is assigned to 
a possible change of the acceleration mechanism of cosmic rays.

Another feature in the spectrum is a small break slightly below $10^{17}\,$eV. 
Applying a second power law above 
$10^{17}\,$eV an index of $\gamma=-3.39 \pm 0.07$ is obtained. 
The indices of the two power-laws differ from each other by two standard deviations. 
Even taking into account extreme scenarios for the systematic uncertainties, or 
applying more stringent procedures to calculate the significance an effect 
with $>1 \sigma$ remains.
Fitting the spectrum with a function of two power laws intercepted by a smooth knee
the energy of the break is assigned to $\log_{10} (E/eV) = 16.92 \pm 0.09$, which is 
in nice agreement to the value obtained by analysing the raw-like 
(i.e.~not corrected for reconstruction uncertainties) all-particle spectrum~\cite{kg-letter}.
In~\cite{kg-letter} it was also seen that the break gets more significant when analysing a 
subsample of events where showers generated by heavy primary particles are enhanced.
The change in slope occurs at an energy where the charge dependent knee 
of the iron component would be expected (KASCADE QGSJet based analysis assigns the proton 
knee to an energy of $\approx 3 \cdot 10^{15}\,$eV). 
The change of the spectral index of this knee-like feature is small compared to the 
first one, original well-known knee~\cite{kas-unf}, what could be explained, 
when the iron component is not dominant around $10^{17}\,$eV. This again can happen
in presence of a `component B' of mixed composition, but a final conclusion is not 
possible without investigating the composition in detail. 

Both observed features were subject to detailed cross-checks. In particular, we investigated 
how far the applied unfolding procedure affects the spectrum. To build up the response matrix
an energy spectrum and a particular composition has to be assumed. We investigated possible effects by 
assuming extreme cases and by using different unfolding methods. 
If one assumes a very abrupt change of the spectral slope and in composition for a given energy, 
the resolution of KASCADE-Grande would indeed smear that out to a structure distributed over values of 
$0.3\,-\,0.5$ in $\log_{10}$(E/eV) of the reconstructed spectrum, but still clearly visible.

At higher energies the KASCADE-Grande spectrum, in particular close to $10^{18}\,$eV, where other 
experiments have claimed a `second knee'~\cite{second}, suffers from missing statistics.    

Despite the fact, that the discussed spectrum is based on the specific hadronic interaction model QGSJet-II, 
there is confidence that the found structures of the energy spectrum remain stable. 
The analysis has shown that the applied procedure can 
reconstruct the total number of charged particles, 
as well as the total muon number sufficiently well, independently of the hadronic interaction model in use.
But the energy calibration assumes that the QGSJet-II model 
provides the correct lateral distribution of the particles over the entire distance range 
(exceeding the geometrical size of KASCADE-Grande). 
First studies with an alternative method to reconstruct 
the energy spectrum via the particle density at a fixed distance give hints to systematic 
deviations~\cite{toma} in the energy calibration of the observable. 
But, the spectral structures discussed above are also present in the results of these studies.
\begin{figure*}[!t]
\vspace*{0.05cm}
\begin{center}
\includegraphics [width=1.0\textwidth]{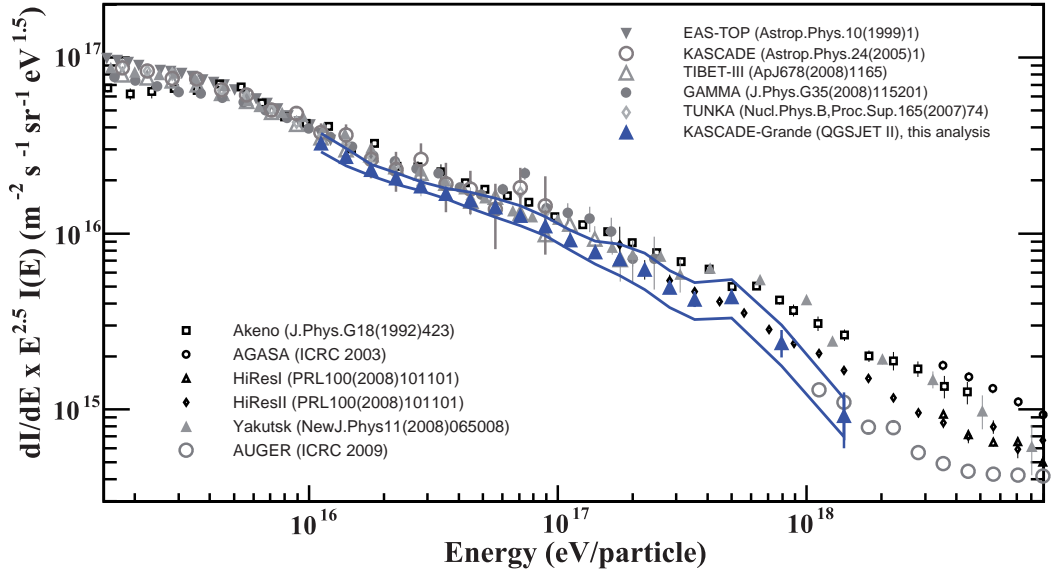}
\end{center}
\vspace*{-0.3cm}
\caption{Comparison of the all-particle energy spectrum obtained with
KASCADE-Grande data based on the QGSJet-II model to results of other experiments. 
The band denotes the systematic uncertainties. An analysis based on EPOS 1.99 would result in a spectrum which is 
shifted downwards by approximately 10\% in intensity.}
\label{spectrum}
\end{figure*}

Figure~\ref{spectrum} compiles the KASCADE-Grande energy spectrum with results of other experiments.
Despite the independent measurements and data analysis there is a good agreement 
with the results of the KASCADE experiment and others in the overlapping energy range at low energies. 
In particular, the concave behavior seems to be needed to connect the spectrum with the spectra 
obtained by other experiments at the knee region. 
At higher energies the KASCADE-Grande spectrum (QGSJet-II) results in a slightly lower intensity
compared to earlier experiments, in particular GAMMA, AKENO and YAKUTSK. 
The strong peak-like structure below $10^{17}\,$eV as was claimed by the GAMMA experiment~\cite{gamma} 
is not confirmed by our results.
At the highest energy accessible by the KASCADE-Grande experiment, where we suffer from missing statistics, 
our result is in agreement with a single power law and with the spectrum reported by HiRes and, 
when taken into account also the systematic 
uncertainties mentioned for the Auger result, with the Pierre Auger Observatory.  
 
\section{Conclusion}

The main air-shower observables of KASCADE-Grande, shower size and total number of muons,
are reconstructed with high precision and low systematic uncertainties. 
Applying various reconstruction methods to the KASCADE-Grande data  
the obtained all-particle energy spectra are compared as a way to cross-check the reconstruction, 
to study systematic uncertainties and to test the validity of the underlying 
hadronic interaction models. 
By combining both observables, the
all-particle energy spectrum of cosmic rays is reconstructed in the energy range of $10^{16}\,$eV to
$10^{18}\,$eV within an uncertainty in intensity of 10-15\%, based on the hadronic interaction model QGSJet-II.  

Correcting the spectra for reconstruction uncertainties and taking into account the systematic 
uncertainties for all methods, the underlying hadronic interaction models (QGSJet-II/FLUKA) result 
in a consistent solution, independent on the observable used, i.e.~the single shower sizes or the 
correlation between the different observables. 
Tests with the hadronic interaction model EPOS 1.99 
have shown that there is a shift in the absolute energy scale when interpreting the data with
this model, but the shape of the spectrum with its structures stays preserved. 
Progress in improving the interaction models is expected in the near future by detailed analyses of 
the now available data of the Large Hadron Collider, LHC (see, e.g.~\cite{pierog}). 
 
The resulting spectrum is consistent, and in the overlapping energy range in a 
very good agreement, with results of the KASCADE, EAS-TOP, and other experiments 
(Fig.~\ref{spectrum}). 
The all-particle energy spectrum in the range from $10^{16}\,$eV to
$10^{18}\,$eV is found to exhibit some smaller structures:
In particular, a hardening of the spectrum is observed at $2\cdot10^{16}\,$eV and a 
small break-off at around $8\cdot10^{16}\,$eV. 
These features are used to discuss the astrophysics in the transition 
region from galactic to extragalactic origin of cosmic rays, where a final conclusion 
is not possible without detailed knowledge of the elemental composition in this energy range. 
However, amongst others, the model proposed by Hillas~\cite{hillas}, e.g., which assumes a second 
component of galactic cosmic rays in addition to the standard SNR component, can explain the 
observed features of the measured all-particle energy spectrum.

A wealth of information on individual showers is available with KASCADE-Grande. 
This makes it possible to reconstruct the all-particle energy spectrum with high precision, 
as well as to investigate the elemental composition, to test hadronic interaction models, 
and to study cosmic ray anisotropies. 
All these studies are under way and further results are expected in the near future.

\begin{ack}
KASCADE-Grande is supported by
the BMBF of Germany, the MIUR and INAF of Italy, the
Polish Ministry of Science and Higher Education (this work partly by grant for 2009-2011).
This work was partially supported by the Romanian Authority for Scientific Research CNCSIS-UEFISCSU 
grant PNII-IDEI no.461/2009, code 1442/2008 and project PN 09 37 01 05, 
and the German-Mexican bilateral collaboration grant (DAAD-Proalmex 2009-2012).
J.C.A. acknowledges the partial support of CONACyT and the Coordinaci\'on de 
la Investigaci\'on Cient\'\i fica de la Universidad Michoacana.
\end{ack}

\begin{appendix}

\section{Appendix}

\subsection{Reconstruction of the total muon number}
\label{appmuon}

\begin{figure}[hbt]
\vspace*{0.05cm}
\begin{center}
\includegraphics [width=0.98\textwidth]{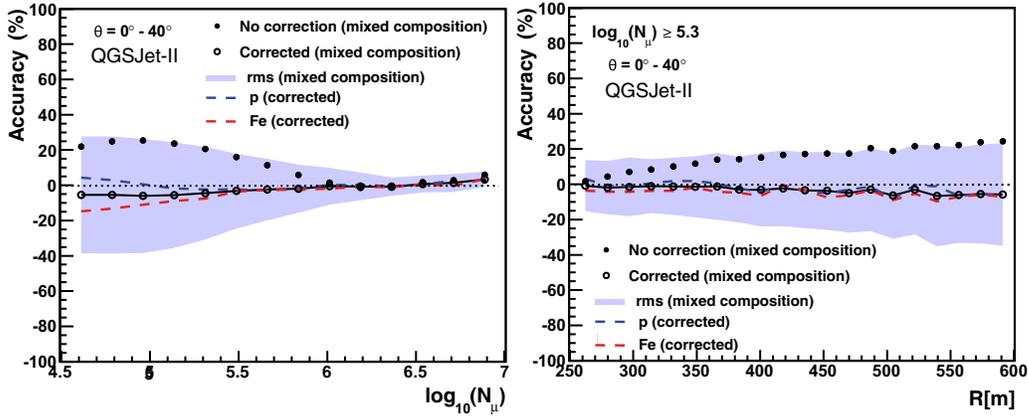}
\end{center}
\vspace*{-0.1cm}
\caption{Accuracy ($\Delta N_\upmu / N_\upmu$ with $\Delta N_\upmu = N^{True}_\upmu - N_\upmu$) 
of the reconstructed muon number before (solid circles) and after 
(open circles) applying the muon correction function versus the 
simulated muon size (left panel) and the distance of the EAS 
core to the KASCADE center (right panel). 
The lines represent the mean values of the achieved $N_\upmu$ accuracy for
different primaries after applying the correction function and the band, the
corresponding one-sigma width for a mixed composition assumption.}
\label{figcorr}
\vspace*{-0.01cm}
\end{figure}

Due to the fact that the muon detectors are located at the fringe of the Grande array, 
the uncertainty of the reconstructed muon number grows is seen to increase with the distance 
of the KASCADE array to the shower core from about 5-10\% at $250\,$m to 25\% 
at $600\,$m for $\log_{10}(N_{\upmu}) > 5.3$.  
But, as the features of these inaccuracies are well understood, we correct the reconstructed 
muon number by a correction function calculated on the basis of Monte Carlo simulations.
This function takes simultaneously into account the dependence of the $N_\mu$ uncertainties on 
the zenith angle of the reconstructed air shower, on the distance of the core position from 
the KASCADE array in shower coordinates, and on the muon number. 
The correction function was found to be nearly independent from the composition of cosmic rays 
and from the hadronic interaction model used.
The uncertainty after applying these corrections is less than 8\% for total 
muon numbers above $\log_{10}(N_{\upmu}) > 5.3$ (Fig.~\ref{figcorr}), 
i.e. in the range of full efficiency. 

\subsection{Response matrix and unfolding}
\label{appunf}

As the fluctuations in the energy determination are larger than the bin size of the aimed-for  
 energy spectrum an unfolding procedure is applied. 
 Using Monte Carlo simulations a response matrix is constructed for the energy 
 interval $\log_{10}(E/\mbox{GeV}) = 6 - 9.5$, i.e.~covering the entire range
 where fluctuations can affect the energy spectrum. 
 This matrix represents the conditional probability, $P(E_j|E^{true}_i)$, of
 an event with true energy in the bin $\log_{10}(E^{true}_i)$ being
 reconstructed with energy $\log_{10}(E_j)$. By means of the response matrix
 a system of simultaneous equations, $n^{exp}_{j} = \sum_{i=1}^N
 P(E_j|E^{true}_i) n^{true}_{i}$, is established between the distribution 
 of measured events, $n^{exp}_{i}$, and the actual energy distribution, 
 $n^{true}_{i}$. The system is solved iteratively for the true distribution
 by means of the Gold-algorithm \cite{gold,ulrich01}. 
 Statistical uncertainties on the data are taken into account during unfolding 
 by multiplying the equation system by the error matrix 
 $C_{ij} = \delta_{ij}/\sigma(n^{exp}_{j})$ \cite{ulrich01}. The unfolding
 procedure involves also a van Cittert transformation of the equation system
 to guarantee the positive definiteness of the modified response matrix,
 which is a necessary requirement for the convergence of the method
 \cite{gold}. On the other hand, unfolded results are validated with another 
 recursive method known as the Bayes-algorithm \cite{agostini}. In general, 
 both unfolding techniques perform in a stable way for the present application. 

 To avoid the problem of having wild fluctuations when increasing the number
 of iterations in the procedure a regularization method is applied, consisting
 of smoothing the result of unfolding in a given step before using it in the 
 next iteration \cite{agostini}. Smoothing was also applied to the response
 matrix, to  avoid the presence of artificial effects in the unfolded
 distribution, which could arise from random fluctuations in some entries of
 the response  matrix as a consequence of the limited statistics of the Monte
 Carlo data sets. 
 Different methods for smoothing are applied and compared in order to 
 find the optimal parameters for the procedure. To smooth
 the spectrum best results are obtained with the $353HQ$-twice
 algorithm\footnote {In each step of the iteration, $k$, the input spectrum, 
 plotted as $\log_{10}[n^{(k-1)}_{i}]$ vs $\log_{10}(E^{true}_i)$, is smoothed
 with the aforementioned algorithm.}\cite{Friedman}. For the response matrix,
 quadratic fits along the diagonals are performed in the region of full 
 efficiency to interpolate data into the region of low statistics. It is worth
 to mention that among the several tests employed to verify the performance of the
 methods, it was checked that the algorithms do not produce artificial structures in 
 the spectrum or hide peaks which could be significantly present in the data,
 besides it was searched for the agreement between the forward-folded and measured 
 distributions.
  
 For a large number of iterations, both the Gold- and Bayes-algorithms
 converge to the same result in the energy interval of full efficiency and good 
 statistics. In addition, positiveness of the solution is observed. 
 As one goes deeper in the number of steps, the weighted mean of the 
 squared sum of statistical errors and systematic bias, i.e
 $(1/N)\sum_{i}^{N}(\sigma^2_{stat,i} + \sigma^2_{syst,i})/n^{true}_{i}$, 
 decreases and becomes constant when the result converges. This parameter
 speaks about the quality of the final solution. Unfolding without smoothing 
 was also tried, but, in general, poorer results were obtained.

\subsection{Attenuation correction}
\label{appatt}

\begin{figure}[t]
\vspace*{0.05cm}
\begin{center}
\includegraphics [width=0.98\textwidth]{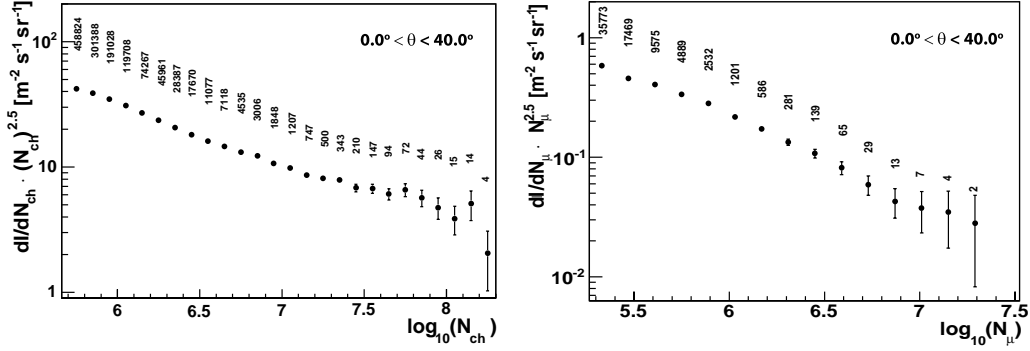}
\end{center}
\vspace*{-0.1cm}
\caption{CIC corrected size spectra with statistical error bars and number of 
events per bin.}
\label{CICspec}
\vspace*{-0.01cm}
\end{figure}

In order to cross-check the attenuation effects of the observables $N_{ch}$ and $N_\upmu$ on the treatment in 
reconstructing the all-particle spectrum, we applied to the two estimated observable spectra
a correction which is independent of Monte Carlo simulations. 
The Constant Intensity Cut Method (CIC method) is based on the assumption that the arrival 
direction distribution of cosmic rays is isotropic and that the cosmic ray intensity and composition 
changes monotonically with primary energy~\cite{HP,Winn,Alva}. 
In this way, the intensity of primary particles becomes a reference variable for 
the primary energy of cosmic rays independent of the zenith angle. 
To apply the CIC method the integral spectra, $J(>N_{ch})$ 
and $J(>N_\mu)$, are calculated  for all angular bins. 
Then, fixed frequency rates (integral intensities) are chosen in the range of maximum 
efficiency and sufficient statistics. 
By this, attenuation curves for each intensity are built, where an 
interpolation between two adjacent points of the integral spectrum is applied. 
The evolution of the shower size (muon number) in the atmosphere is extracted 
from the attenuation curves. 
The data have shown that it is possible to use constant attenuation parameters for the 
entire energy range since the differences in 
the obtained parameters by fitting individual curves are smaller than the uncertainty. 
With the parameters obtained, the shower size (muon number) 
of an individual air shower can be corrected with  
$N_{ch(\upmu)}(\theta_{ref}) = N_{ch(\upmu)} (\theta) \exp \left[ P(\theta_{ref}) - P(\theta) \right]$ 
to obtain the equivalent size at a given zenith angle of reference, $\theta_{ref}$.
The reference angle is chosen to be the mean of the measured zenith angle distribution, 
which is found to be $20^\circ$ and $22^\circ$ for the shower size and muon number 
distributions, respectively. 
Due to the independent reconstruction of $N_{ch}$ and $N_{\upmu}$ slightly 
different reconstruction thresholds lead to different mean angles and hence, different 
reference angles~\cite{kang,arteaga}.
Uncertainties due to intrinsic assumptions on 
energy independent shower-to-shower fluctuations, the assumed spectral index, and 
constant composition are estimated and are taken into account.
In Fig.~\ref{CICspec} the resulting spectra as corrected by the CIC 
method are shown for the shower size and for the muon number in the range of full 
efficiency for the independent reconstruction of the observables. 
Note that the bin sizes are selected according to the resolution of $N_{ch}$ and $N_{\upmu}$.
Applying this procedure to simulated data (QGSJet-II) has shown that indeed 
the differences between simulation and data are small for showers below 40$^o$ and that 
the estimated uncertainty on our all-particle spectrum is a conservative approximation.

\subsection{Single parameter reconstruction of the energy spectrum}
\label{apponed}

To reconstruct the energy spectrum by the observables independently, we use 
the reconstructed shower size spectra given in Fig.~\ref{figrawspecs} and apply first an 
attenuation correction to the observables.  
Then, the shower size per individual event is calibrated by Monte Carlo simulations 
under the assumption of correlations in the form $E_0\propto N_{ch}^{\alpha_{ch}}$  and 
$E_0\propto N_{\upmu}^{\alpha_\upmu}$, respectively, and an assumed 
primary composition~\cite{kang,arteaga}. 

\begin{figure}[t]
\vspace*{0.05cm}
\begin{center}
\includegraphics [width=0.98\textwidth]{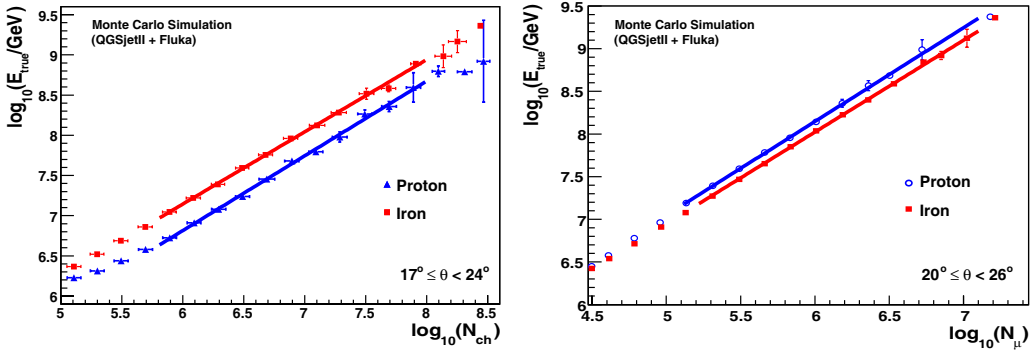}
\end{center}
\vspace*{-0.1cm}
\caption{
Calibration functions for assumed pure proton and iron primaries for the 
observables $N_{ch}$ (left panel) and $N_\upmu$ (right panel), respectively.}
\label{calib}
\vspace*{-0.01cm}
\end{figure}

To determine the calibration function of the number of 
(attenuation corrected) charged particles $N_{ch}$ and primary energy, Monte Carlo 
simulations were used, where a zenith angle range of $17^\circ \leq \theta 
< 24^\circ$, i.e.~around the reference angle, was selected. 
Assuming a linear dependence in logarithmic scale: $\log_{10}{E} = a + b \cdot \log_{10}{N_{ch}}$, 
the correlation between the primary energy and the number of charged particles is obtained,
where the fit is applied in the range of full trigger and reconstruction efficiencies
(see left panel of Fig.~\ref{calib}).  
The fit yields $a = 1.23$ and $b = 0.93$ for primary protons and 
$a = 1.75$ and $b = 0.90$ for iron primaries. 
The numbers confirm the assumption of a power law and show that the slopes of these power 
laws are very similar. However, the normalization parameters assigning the energy are very different 
for the two primaries.  
The fits are also performed for helium, carbon, silicon, and for 
a uniform mixed composition to examine the dependence of the calibration on the assumed 
primary particle type, where the values of the fit parameters are found to be in between 
the above values. 
The energy resolution is estimated from the difference between true and reconstructed
energy, resulting in 32\% and 18\% at $E = 10^{17}\,$eV for proton and iron 
primaries, respectively, with an energy dependence of approximately $1/\sqrt{E}$. 
\begin{figure}[t]
\vspace*{0.05cm}
\begin{center}
\includegraphics [width=0.98\textwidth]{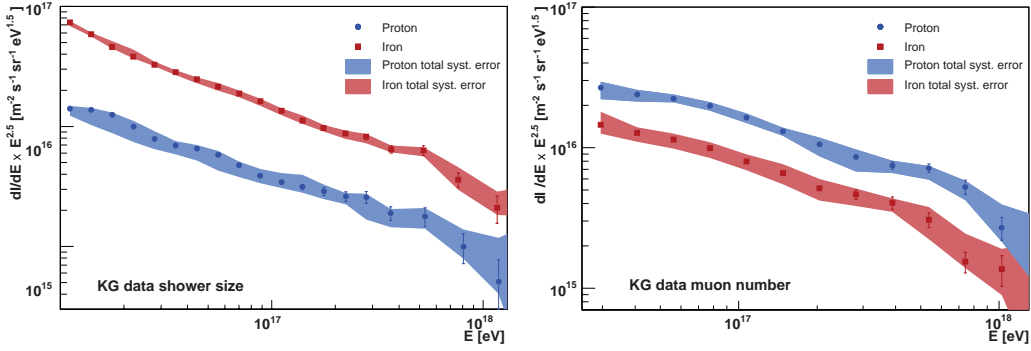}
\end{center}
\vspace*{-0.3cm}
\caption{All-particle energy spectrum (including systematic uncertainties) 
for the assumption of primary protons and iron, respectively, based on the 
shower size (left panel) and on the total muon number (right panel).}
\label{specs1dim}
\vspace*{0.01cm}
\end{figure}

To determine the energy from the (attenuation corrected) total muon number $N_{\upmu}$ 
a similar procedure as in the case of $N_{ch}$ was applied: a calibration function for the 
muon size ($\theta_{ref} = 22^\circ$) in terms of the primary energy was invoked. 
As before, the calibration curve is described with a relation of the form 
$\log_{10}{E} = a_\upmu + b_\upmu \cdot \log_{10}{N_{\upmu}}$ (see Fig.~\ref{calib}, right panel). 
For the case of pure protons, the fit results in $a_\upmu = 1.61$ and  
$b_\upmu = 1.09$ and for iron primaries the values
$a_\upmu = 1.63$ and  $b_\upmu = 1.07$  are obtained. 
Again it is observed that the slopes are similar, but here the difference in the parameter 
$a$ is much smaller than in case of the charged particles.
In case of the muon shower size, as the fluctuations are smaller, the energy resolution for 
protons and iron nuclei are of the order of 25\% and 12\% ($E \approx 10^{17}\,$eV), respectively.

Figure~\ref{specs1dim} shows the all-particle energy spectra as obtained after applying the 
calibration functions, as well as the corrections for the bin-to-bin fluctuations for 
both approaches. Apart from the statistical uncertainties also bands are shown 
describing the systematic uncertainties on the intensity, which were determined for 
the two approaches and different primaries independently. 
The considered sources of such uncertainties include the estimate of the calibration functions,
the chosen reference angle for the calibration, the muon number correction function, and the 
application of the unfolding procedures. 
The total systematic uncertainty (i.e.~sum in quadrature of all terms) on the intensity for 
proton and iron is 21\% (7\%) and 10\% (13\%), respectively in case of $N_{ch}$ ($N_\upmu$),
at energies of $10^{17}\,$eV.
The systematic uncertainties are energy dependent and evolve in such a way that 
they slightly increase near the threshold, where fluctuations are larger, and in the 
high-energy region, where statistics decreases.

\end{appendix}

\end{document}